\documentclass[aps,prd,superscriptaddress,nofootinbib,tighten,preprint]{revtex4}
\usepackage{amsfonts}
\usepackage{amssymb}
\usepackage{amsmath,color}
\usepackage{epsfig}
\usepackage{graphicx,epsf,epsfig,subfigure}
\usepackage{bbm}

\begin{document}

\preprint{SISSA 29/2014/FISI}
\preprint{RM3-TH/14-9}
\preprint{NORDITA-2014-64}

\title{Constraining Sterile Neutrinos Using Reactor Neutrino Experiments}

\author{Ivan Girardi}
\email{igirardi@sissa.it}

\affiliation{SISSA/INFN, Via Bonomea 265, 34136
Trieste, Italy}

\author{Davide Meloni}

\email{meloni@fis.uniroma3.it}

\affiliation{Dipartimento di Matematica e Fisica,
Universit\`{a} di Roma Tre, Via della Vasca Navale 84, 00146 Rome, Italy}

\author{Tommy Ohlsson}

\email{tohlsson@kth.se}

\affiliation{Department of Theoretical Physics, School of
Engineering Sciences, KTH Royal Institute of Technology, AlbaNova
University Center, 106 91 Stockholm, Sweden}

\author{He Zhang}

\email{he.zhang@mpi-hd.mpg.de}

\affiliation{Max-Planck-Institut f\"{u}r Kernphysik, Saupfercheckweg
1, 69117 Heidelberg, Germany}

\author{Shun Zhou}
\email{shunzhou@kth.se}

\affiliation{Department of Theoretical Physics, School of
Engineering Sciences, KTH Royal Institute of Technology, AlbaNova
University Center, 106 91 Stockholm, Sweden}

\begin{abstract}
Models of neutrino mixing involving one or more sterile neutrinos have resurrected their importance in the light of recent cosmological data. In this case, reactor antineutrino experiments offer an ideal place to look for signatures of sterile neutrinos due to their impact on neutrino flavor transitions. In this work, we show that the high-precision data of the Daya Bay experi\-ment constrain the 3+1 neutrino scenario imposing upper bounds on the relevant active-sterile mixing angle $\sin^2 2 \theta_{14} \lesssim 0.06$ at 3$\sigma$ confidence level for the mass-squared difference $\Delta m^2_{41}$ in the range $(10^{-3},10^{-1}) \, {\rm eV^2}$. The latter bound can be improved by six years of running of the JUNO experiment, $\sin^22\theta_{14} \lesssim 0.016$,  although in the smaller mass range $ \Delta m^2_{41} \in (10^{-4} ,10^{-3}) \, {\rm eV}^2$. We have also investigated the impact of sterile neutrinos on precision measurements of the standard neutrino oscillation parameters $\theta_{13}$ and $\Delta m^2_{31}$ (at Daya Bay and JUNO), $\theta_{12}$ and $\Delta m^2_{21}$ (at JUNO), and most importantly, the neutrino mass hierarchy (at JUNO).  We find that, except for the obvious situation where  $\Delta m^2_{41}\sim \Delta m^2_{31}$,  sterile states do not affect these measurements substantially.
\end{abstract}

\maketitle

\section{introduction}
\label{sec:intro}

Neutrino physics has entered the phase of precision measurements. With the upcoming data in the future, the focus is now to determine the missing fundamental parameters such as the neutrino mass hierarchy, the leptonic Dirac CP-violating phase, and the absolute neutrino mass scale. In addition to the standard neutrino parameters, theoretical and phenomenological investigations of beyond-the-Standard-Model scenarios as sub-leading effects have therefore full attention. Such scenarios include non-standard neutrino interactions, unitarity violation, CPT and Lorentz invariance violation, and models with sterile neutrinos. In this work, we will investigate the impact of sterile neutrinos on the fundamental neutrino parameters and how to constrain sterile neutrinos using reactor neutrino experiments such as the ongoing Daya Bay and upcoming JUNO experiments.

Sterile neutrinos, or strictly speaking fermionic SM singlets, are even more elusive than ordinary active neutrinos, since they are supposed to interact through the gravitational force only, and not through the weak force as active neutrinos. However, sterile neutrinos could mix with active neutrinos, which calls for physics beyond the Standard Model. At the moment, there are three experimental results from neutrino oscillation experiments, which give hints that sterile neutrinos could exist. These three results, usually referred to as anomalies, are the LSND (and MiniBooNE) anomaly \cite{Aguilar:2001ty,AguilarArevalo:2008rc,AguilarArevalo:2010wv}, the Gallium anomaly \cite{Giunti:2006bj,Giunti:2010zu,Giunti:2012tn}, and the reactor anomaly \cite{Mention:2011rk}, which all point to sterile neutrinos with mass of the order of 1~eV and small mixing. It should be noted that if such sterile neutrinos exist, they could be produced in the early Universe, and have played an important role in the cosmological evolution. Global fits to data from short-baseline neutrino oscillation experiments suggest that the data can be described by either three active and one sterile (3+1) neutrinos or three active and two sterile (3+2) neutrinos \cite{Kopp:2013vaa}. However significant constraints come from experiments which would appear to disfavor these anomalies \cite{Armbruster:2002mp,Astier:2003gs}. Even in the case in which these anomalies will be explained not by the presence of sterile neutrinos, there are strong indications on the possibility of the existence of sterile neutrinos with masses lower than $1~{\rm eV}$. Moreover, in viewing of the recent detection of B mode polarization from the BICEP2 experiment \cite{Ade:2014xna}, an analysis of the combined CMB data in the framework of LCDM+r models gives $N_{\rm eff} = 4.00 \pm 0.41$ \cite{Giusarma:2014zza}, in favor of the existence of extra radiation. In this work, we will concentrate on the scenario with 3+1 neutrinos analyzing the constraints from reactor neutrino experiments in a wide range of sterile neutrino masses.

From a theoretical point of view, a lot of novel models have been constructed with the aim of embedding sterile neutrinos in a more fundamental framework. Such possibilities include models of extra dimensions with exponentially suppressed sterile neutrino masses, see for instance Ref.~\cite{Kusenko:2010ik,Adulpravitchai:2011rq}. A slightly-breaking flavor symmetry model may generate a neutrino with much smaller mass than the other two, whose masses are allowed by the symmetry. This has been proposed to generate seesaw neutrinos of keV scale in Refs.~\cite{Shaposhnikov:2006nn,Lindner:2010wr}, see also Ref.~\cite{Mohapatra:2005wk}. While the commonly studied flavor models with non-Abelian discrete symmetries cannot easily produce a non-trivial hierarchy between fermion masses, the Froggatt--Nielsen mechanism is capable of such a production~\cite{Froggatt:1978nt}. This has been proposed to generate seesaw neutrinos of eV--keV scales in Refs.~\cite{Barry:2011fp,Merle:2011yv,Barry:2011wb}. Extensions or variants of the canonical type-I seesaw mechanism often contain additional mass scales, which can be arranged to generate light sterile neutrinos~\cite{Barry:2011wb,Zhang:2011vh,Heeck:2012bz,Dev:2012bd}.

The latest measurements of the small mixing angle $\theta_{13}$ have been established by several neutrino oscillation experiments, but it was the Daya Bay experiment that first found a statistical significance of more than 5$\sigma$ confidence level, and therefore, won the hunt for this mixing angle \cite{An:2012eh}. Apart from the standard oscillation picture, the reactor antineutrino experiments could also help us to probe new physics as non-standard effects in neutrino oscillations~\cite{Ohlsson:2008gx,Leitner:2011aa,Adhikari:2012vc,Khan:2013hva,Girardi:2014gna,Bakhti:2014pva}. In this work, we will use the existing data of the Daya Bay experiment as well as the sensitivity of the future JUNO experiment to put constraints on sterile neutrinos using scenarios with 3+1 neutrinos \cite{Palazzo:2013bsa,Bakhti:2013ora,Esmaili:2013yea}.

This work is organized as follows: In Sec.~\ref{sec:neu_osc_prob}, we will analyze neutrino oscillation probabilities with three active neutrinos and one sterile neutrino both (i) analytically and (ii) numerically. Especially, we will consider three cases for the probabilities based on different regimes of the neutrino mass-squared differences, and the effects of sterile neutrinos on the determination of neutrino mass hierarchy. Then, in Sec.~\ref{sec:DB}, we will investigate the impact and signature of one sterile neutrino using the existing data from the Daya Bay experiment \cite{An:2013uza}. Next, in Sec.~\ref{sec:impact_prec}, we will study the sensitivity to sterile neutrino parameters as well as the impact of one sterile neutrino on the precision measurement of the standard neutrino parameters at the JUNO experiment. Finally, in Sec.~\ref{sec:s&c}, we will summarize the results and present our conclusions.

\section{Neutrino Oscillation Probabilities}
\label{sec:neu_osc_prob}

In the presence of $n$ sterile neutrinos, the neutrino mass matrix is an $(n+3) \times (n+3)$ matrix, which can be diagonalized by means of an $(n+3) \times (n+3)$ unitary matrix $U$. In general, one has $(n+3)(n+2)/2$ mixing angles and $(n+2)(n+1)/2$ Dirac phases. In the case of only one sterile neutrino, $U$ is typically parameterized by
\begin{equation}
U = R_{34}\tilde{R}_{24}\tilde{R}_{14}R_{23}\tilde{R}_{13}R_{12}P \,
, \label{eq:UPMNS4x4}
\end{equation}
where the matrix $R_{ij}$ is a rotation by the angle $\theta_{ij}$ in the corresponding $ij$ space, e.g.
\begin{equation}
R_{34} = \begin{pmatrix} 1 & 0 & 0 & 0 \\ 0 & 1 & 0 & 0 \\ 0 & 0 &
c_{34} & s_{34} \\ 0 & 0 & -s_{34} & c_{34} \end{pmatrix} \quad {\rm
or} \quad \tilde{R}_{14} = \begin{pmatrix} c_{14} & 0 & 0 &
s_{14}e^{-i\delta_{14}} \\ 0 & 1 & 0 & 0 \\ 0 & 0 & 1 & 0 \\
-s_{14}e^{i\delta_{14}} & 0 & 0 & c_{14} \end{pmatrix}
\end{equation}
with $s_{ij} = \sin\theta_{ij}$ and $c_{ij}=\cos\theta_{ij}$.
The diagonal matrix $P$ contains three Majorana phases, which are irrelevant to our discussion. In this parametrization, one can figure out that
\begin{equation}
\left| U_{e1} \right| = c_{14} c_{13} c_{12} \; , ~~~ \left| U_{e2} \right|  = c_{14} c_{13} s_{12} \; , ~~~ \left| U_{e3} \right| = c_{14} s_{13} \; , ~~~ \left| U_{e4} \right| = s_{14} \; ,
\label{eq:Uei}
\end{equation}
indicating that only the mixing angle $\theta_{14}$ enters reactor electron antineutrino oscillations.

The survival probability of electron antineutrinos from nuclear reactors can be written as
\begin{equation}
P_{\bar{e}\bar{e}} \equiv P_{ee} = 1 - 4\sum_{i < j}|U_{e i}|^2 |U_{e j}|^2 \sin^2 \Delta_{ji} \; ,
\label{eq:prob}
\end{equation}
where $\Delta_{ji} \equiv \Delta m^2_{ji} L/(4E)$ denote the oscillation phases, $L$ the baseline length, $E$ the neutrino energy, and $\Delta m^2_{ji} \equiv m^2_j - m^2_i$ the mass-squared difference of two neutrino mass eigenstates $i$ and $j$. Using Eq.~\eqref{eq:Uei}, we can rewrite the survival probability~\eqref{eq:prob} as
\begin{eqnarray}
P_{ee} &=& 1 - c^4_{14} s^2_{12} \sin^2 2\theta_{13} \sin^2 \Delta_{32} - c^4_{14} c^2_{12} \sin^2 2\theta_{13} \sin^2 \Delta_{31}   - c^4_{14} c^4_{13} \sin^2 2\theta_{12} \sin^2 \Delta_{21} \nonumber \\
&& - s^2_{13} \sin^2 2\theta_{14} \sin^2 \Delta_{43} - c^2_{13} s^2_{12} \sin^2 2\theta_{14} \sin^2 \Delta_{42} - c^2_{13} c^2_{12} \sin^2 2\theta_{14} \sin^2 \Delta_{41}  \; ,
\label{eq:probexp}
\end{eqnarray}
where the oscillation terms are cast into two rows. The first row collects the contributions from active neutrinos, while the second row from sterile neutrinos.
In what follows, we will concentrate on the oscillation probability at the Daya Bay and JUNO setups, in which one of
the standard oscillation modes in the first row of Eq.~\eqref{eq:probexp} dominates the probability.
In addition, the $\Delta_{43}$ mode is further suppressed by both $\theta_{13}$ and $\theta_{14}$, and can therefore be safely neglected.
Thus, the oscillation probability~\eqref{eq:probexp},
in the limit $c^2_{13} = c^2_{14} = 1$, approximates to
\begin{eqnarray}
P_{ee} & \simeq & 1 -  s^2_{12} \sin^2 2\theta_{13} \sin^2 \Delta_{32} -  c^2_{12} \sin^2 2\theta_{13} \sin^2 \Delta_{31}   -   \sin^2 2\theta_{12} \sin^2 \Delta_{21} \nonumber \\
&&  - s^2_{12} \sin^2 2\theta_{14} \sin^2 \Delta_{42} -  c^2_{12} \sin^2 2\theta_{14} \sin^2 \Delta_{41}  \; .
\label{eq:probapp}
\end{eqnarray}

\subsection{The Electron Antineutrino Survival Probability at Daya Bay}

Since the baseline length of the Daya Bay detectors is relatively short, the $\Delta_{21}$ related modes are strongly suppressed by $L$, and it is a good approximation to use $\Delta_{32} \simeq \Delta_{31}$ and $\Delta_{42} \simeq \Delta_{41}$. Hence, the oscillation probability~\eqref{eq:probapp} is simplified to
\begin{eqnarray}
P_{ee} \simeq 1 -  \sin^2 2\theta_{13}  \sin^2 \Delta_{31}  -    \sin^2 2\theta_{14}  \sin^2 \Delta_{41} \; ,
\end{eqnarray}
where terms like $s^2_{13} s^2_{14}$ have been dropped. The last term appears in short-baseline reactor neutrino experiments when $|\Delta m^2_{41}| \gtrsim 10^{-3}~{\rm eV}^2$, and may play an important role in explaining the reactor neutrino anomaly. Furthermore, the sterile neutrino contributions would make significant modifications to the electron antineutrino spectrum. In case of a larger active-sterile mass-squared difference, the second term leads to fast oscillations, which result in a shift of the total observed events.

In the interesting situation that $\Delta m^2_{31} \simeq \Delta m^2_{41}$, the two terms in the oscillation probability can be combined, and one can define an effective mixing angle as $\sin^22\tilde \theta_{13} = \sin^22\theta_{13} +\sin^22\theta_{14}$. In this case, sterile neutrinos induce mimicking effects that add a correction to the observed mixing angle $\theta_{13}$. Accordingly, Daya Bay loses its sensitivity to sterile neutrinos.

\begin{figure}[!t]
\begin{center}
\subfigure{%
\hspace{-1.4cm}
\includegraphics[width=0.64\textwidth]{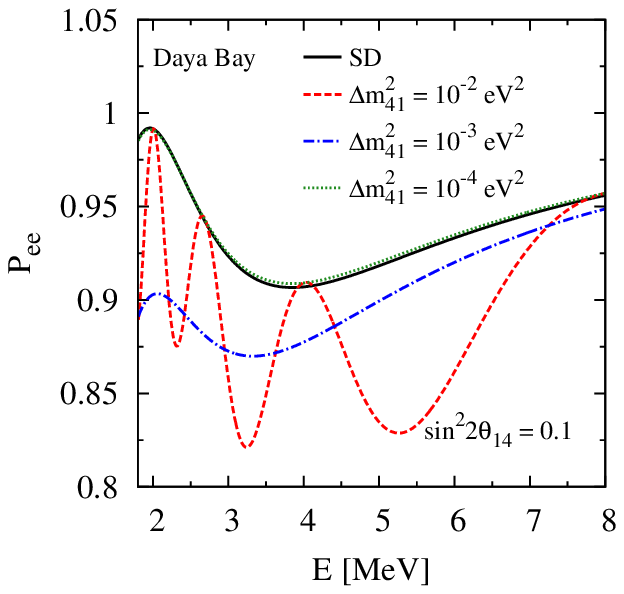}        }%
\subfigure{%
\hspace{-2.4cm}
\includegraphics[width=0.64\textwidth]{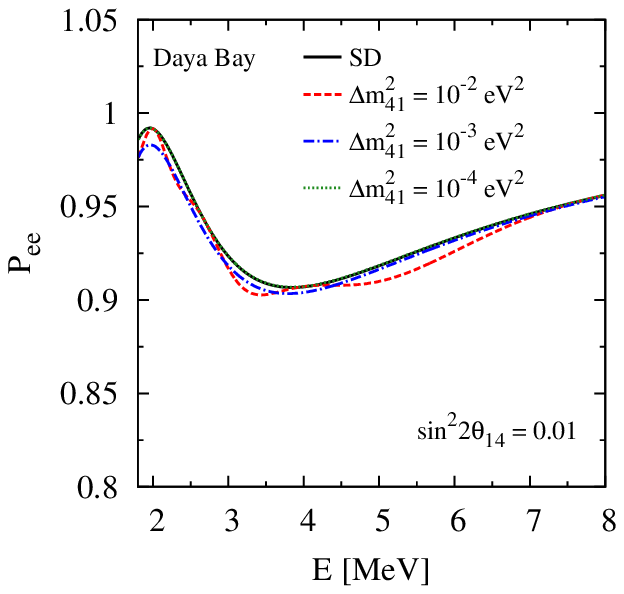}        }
\vspace{-1.2cm}
\caption{\it The oscillation probability $P_{ee}$ at Daya Bay as a function of neutrino energy $E$ for $L=2~{\rm km}$ and $\sin^22\theta_{14}=0.1$ (left plot) as well as $\sin^22\theta_{14}=0.01$ (right plot).
SD refers to the standard oscillation probability.}
\label{fig:PeeDB}
\end{center}
\end{figure}
In Fig.~\ref{fig:PeeDB}, we illustrate the oscillation probability at the Daya Bay far detector with baseline length $L=2~{\rm km}$ and mixing $\sin^22\theta_{14}=0.1$ [cf.~Eq.~\eqref{eq:input} for the other standard oscillation parameters]. As one can read off from the plot, in the limit $|\Delta m^2_{41}| \ll |\Delta m^2_{32}|$, the black solid and green dotted curves almost overlap, and hence, Daya Bay has no sensitivity to sterile neutrinos in this mass regime. In the limit $\Delta m^2_{41} \sim \Delta m^2_{31}$, the sterile polluted curve differs from the standard one. However, this difference can be compensated by taking a smaller value for $\theta_{13}$. A combined analysis of reactor and long-baseline experiments are therefore needed to discriminate this ambiguity. In the regime $|\Delta m^2_{41}| \gg |\Delta m^2_{32}|$, the fast oscillations induced by sterile neutrinos lead to a clear distinction to the standard oscillation behavior, and can be well constrained using the current Daya Bay data.

\subsection{The Electron Antineutrino Survival Probability at JUNO}

Different from the Daya Bay setup, the JUNO detector will be located around 50~km away from the nuclear power plant, indicating that the $\Delta_{21}$ oscillation mode is dominating, whereas the $\Delta_{31}$ and $\Delta_{32}$ related oscillation modes become fast oscillations. The sterile neutrino related oscillation modes $\Delta_{4i}$ induce corrections to the neutrino spectrum. Since the JUNO energy resolution is optimized for the determination of the neutrino mass hierarchy, the JUNO detector turns out to be sensitive to mass-squared differences between $10^{-5}~{\rm eV}^2$ and $10^{-2}~{\rm eV}^2$. Above this mass range, the oscillation frequency is too fast to be distinguished, whereas, below this range, the oscillation behavior does not manifest due to the suppression of baseline length and neutrino energy. Therefore, one may consider the following three cases:
\begin{enumerate}

\item The sterile neutrino is nearly degenerate with one of the three active neutrinos, i.e.~$|\Delta m^2_{4i}| < 10^{-5}~{\rm eV}^2 $ (for $i=1,2$, or $3$). The active-sterile mass-squared differences can be ignored in this case, and the $\Delta_{42}$ and $\Delta_{41}$ terms in Eq.~\eqref{eq:probapp} can always be absorbed into the standard oscillation terms. The role of sterile neutrinos is simply to correct the standard neutrino mixing angles, implying loss of sensitivity to sterile neutrinos.

\item In the case of a much larger active-sterile mass-squared difference, i.e.~$|\Delta m^2_{4i}| > 10^{-2}~{\rm eV}^2 $, the fast active-sterile oscillations are actually beyond the resolution limit of JUNO. In this regime, the Daya Bay setup performs a better probe of sterile neutrinos. The reason is that the baseline length of Daya Bay is much shorter than that of JUNO (the Daya Bay baseline is only about 2~\% of the JUNO baseline), and hence, the fast oscillations at Daya Bay is milder, which provides us with a better chance to distinguish the sterile neutrino induced oscillations from the standard ones.

\item In the range $10^{-5}~{\rm eV}^2 < |\Delta m^2_{4i}| <  10^{-2}~{\rm eV}^2$, the observed neutrino spectrum obtains corrections from sterile neutrinos and one would expect a better sensitivity at JUNO.

\end{enumerate}
\begin{figure}[!t]
\begin{center}
\subfigure{%
\hspace{-1.4cm}
\includegraphics[width=0.64\textwidth]{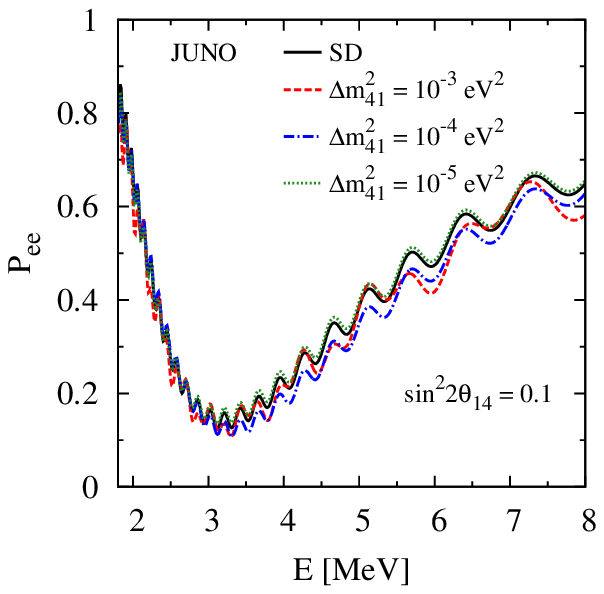}        }%
\subfigure{%
\hspace{-2.4cm}
\includegraphics[width=0.64\textwidth]{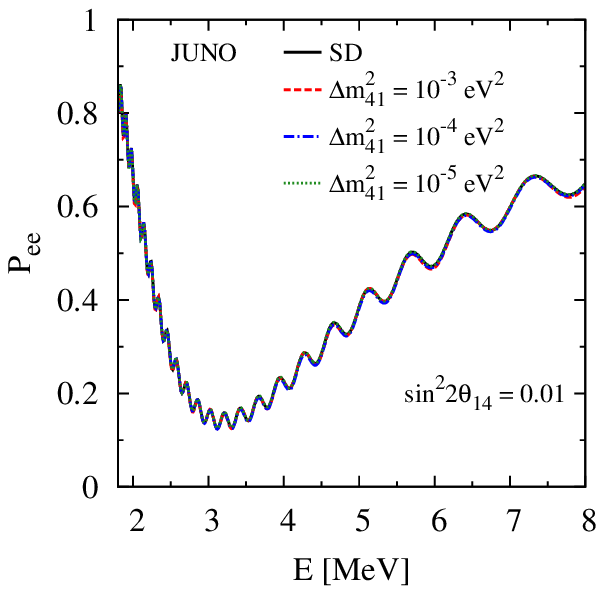}        }
\vspace{-1.2cm}
\caption{\it The oscillation probability $P_{ee}$ at JUNO as a function of neutrino energy $E$ for $L=52.5~{\rm km}$ and $\sin^22\theta_{14}=0.1$ (left plot) as well as $\sin^22\theta_{14}=0.01$ (right plot). Here the normal mass hierarchy is assumed.
SD refers to the standard oscillation probability.}
\label{fig:PeeJUNO}
\end{center}
\end{figure}

In Fig.~\ref{fig:PeeJUNO}, the sterile neutrino corrections are illustrated for the JUNO setup. When the active-sterile mixing is sizable, the effects of sterile neutrinos become more significant in the large energy regime, in particular for the cases $\Delta m^2_{41} = 10^{-3} ~{\rm eV}^2$ and $\Delta m^2_{41} = 10^{-4} ~{\rm eV}^2$. The shift of the energy spectrum provides us with the possibility to search for sterile neutrinos. For the case of a small value for $\theta_{14}$, the deviation from the standard oscillations is less pronounced, and one needs in principle a challenging experimental setup with a very high precision to detect sterile neutrinos.

Since the major purpose of JUNO is to settle the neutrino mass hierarchy, one may wonder if the presence of sterile neutrinos may affect the determination of the neutrino mass hierarchy at JUNO. To this end, we present the probability difference between the normal and inverted mass hierarchy cases:
\begin{eqnarray}
\Delta P & = & P^{\rm NH}_{ee} - P^{\rm IH}_{ee}  \nonumber \\
   & \simeq & 2  \sin2\Delta_{21} \left( s^2_{12} \sin^22\theta_{13}\cos\Delta_{31}\sin\Delta_{31} - c^2_{12} \sin^22\theta_{14}\cos\Delta_{42}\sin\Delta_{42} \right)  \; ,
\label{eq:deltaP}
\end{eqnarray}
where NH stands for the normal mass hierarchy ($m_3 > m_1$) and IH the inverted mass hierarchy ($m_3 < m_1$). One can clearly observe from Eq.~\eqref{eq:deltaP} that there exists a very interesting situation that in the limit
\begin{eqnarray}
\Delta_{42} & \simeq & \Delta_{31} \; , \label{eq:condition0}\\
s^2_{12} \sin^22\theta_{13} & \simeq & c^2_{12} \sin^22\theta_{14} \; ,
\label{eq:condition}
\end{eqnarray}
the probability difference is equal to zero, i.e.~$\Delta P=0$. In this special case, both normal and inverted mass hierarchy fits would give the same minimal $\chi^2$, and the JUNO setup loses its ability to determine the neutrino mass hierarchy. In other words, if JUNO cannot discriminate between its normal and inverted mass hierarchy analyses, a light sterile neutrino with mass of the order $\Delta m^2_{41} \simeq \Delta m^2_{32}$ and mixing $\sin^22\theta_{14} \simeq 0.04$ could then be the underlying reason.
\begin{figure}[]
\begin{center}
\subfigure{%
\hspace{-1.4cm}
\includegraphics[width=0.64\textwidth]{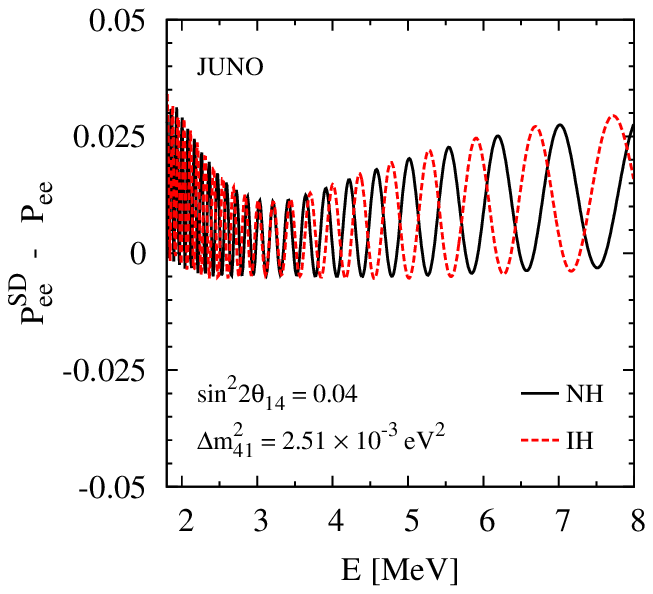}        }%
\subfigure{%
\hspace{-2.4cm}
\includegraphics[width=0.64\textwidth]{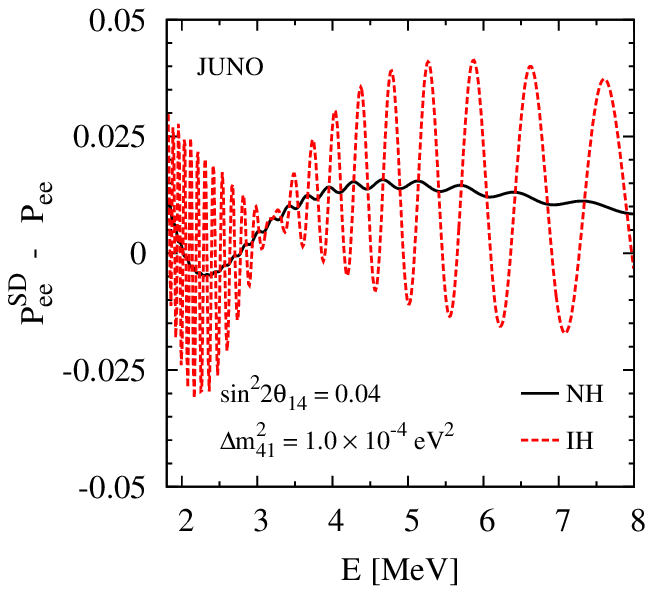}        }
\vspace{-1.2cm}
\caption{\it The probability differences $(P^{\rm SD}_{ee})^{\rm NH} - P_{ee}$ (solid curves) and $(P^{\rm SD}_{ee})^{\rm IH} - P_{ee}$ (dotted curves) as functions of neutrino energy $E$ for $\sin^22\theta_{14}=0.04$ and $\Delta m^2_{42} = \Delta m^2_{31}$ (left plot) as well as $\Delta m^2_{42} = 10^{-4} ~{\rm eV}^2$ (right plot), where $P^{\rm SD}_{ee}$ is the standard neutrino oscillation probability. Here the normal mass hierarchy for $P_{ee}$ is assumed.}
\label{fig:PeeHier}
\end{center}
\end{figure}

In Fig.~\ref{fig:PeeHier}, the impact of sterile neutrinos on the mass hierarchy determination is shown. One can observe from the left plot that when the conditions given in Eq.~\eqref{eq:condition} are fulfilled, normal and inverted mass hierarchy fits will give equally good or equally bad fits to experimental data. In contrast, in the general case, the wrong-hierarchy oscillation probability gives a worse fit, which is clearly seen in the right plot of Fig.~\ref{fig:PeeHier}.

\section{Fit to Daya Bay data}
\label{sec:DB}

In this section, we present the relevant features of the Daya Bay experiment and some of the details of our
statistical analysis.
The Daya Bay experimental setup that we take into account consists of six reactors~\cite{An:2013uza}, emitting
antineutrinos $\bar \nu_e$ whose spectra have been recently
estimated in Refs.~\cite{Mueller:2011nm,Huber:2011wv}.
The total flux of arriving $\bar \nu_e$ at the six antineutrino detectors has been estimated
using the convenient parametrization discussed in Ref.~\cite{Mueller:2011nm} and taking into account
all the distances between the detectors and the reactors (summarised in Tab.~2 of Ref.~\cite{An:2013uza}).
For this analysis we use the data set accumulated during 217 days,
which are extracted from Fig.~2 of Ref.~\cite{An:2013zwz}. The antineutrino energy $E$ is reconstructed
by the prompt energy deposited by the positron $E_{ \rm prompt}$ using
the approximated relation \cite{An:2013uza}
$E \simeq E_{\rm prompt} + 0.8 \; {\rm MeV}$.
The energy resolution
function is a Gaussian function, parametrized according to
\begin{eqnarray}
\sigma(E) [\rm MeV]=
\begin{cases}
\gamma \sqrt{E/\rm MeV - 0.8} \, , \; \mbox{for } E > 1.8 \; \rm MeV \, ,\\
\gamma \, , \; \mbox{for } E \leq 1.8 \; \rm MeV\,,\
\end{cases}
\end{eqnarray}
with $\gamma = 0.08$ MeV. The antineutrino cross section for the inverse beta decay process has been taken from Ref.~\cite{Vogel:1999zy}.

The statistical analysis is performed using a modified version of the GLoBES software \cite{Huber:2007ji,Huber:2003pm,Huber:2004ka} and a $\chi^2$ function which takes into account several sources of systematic errors and retrace the one used by the Daya Bay collaboration.
Details can be found in Ref.~\cite{Girardi:2014gna}.
We analyze the sensitivity of the Daya Bay experiment on the sterile parameters and the
effect of $\theta_{14}$ and $\Delta m^2_{41}$ on the determination of $\theta_{13}$ and $\Delta m^2_{31}$. Fit results have been obtained after a marginalization over the parameters that are not shown in the figures.

\begin{figure}[]
\begin{center}
\subfigure{%
\hspace{-1.4cm}
\includegraphics[width=0.64\textwidth]{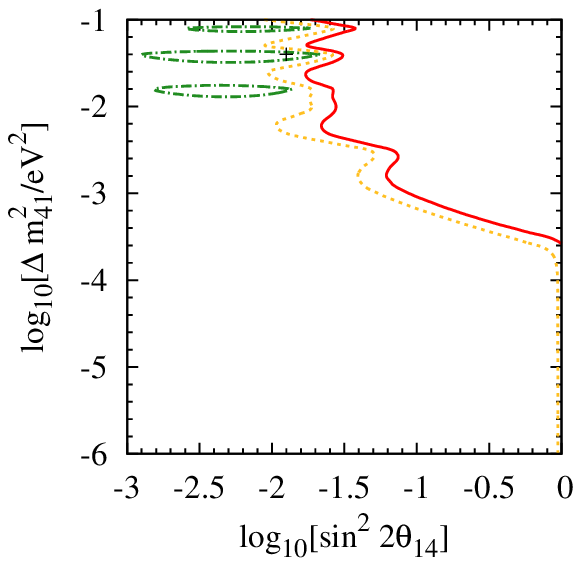}        }%
\subfigure{%
\hspace{-2.4cm}
\includegraphics[width=0.64\textwidth]{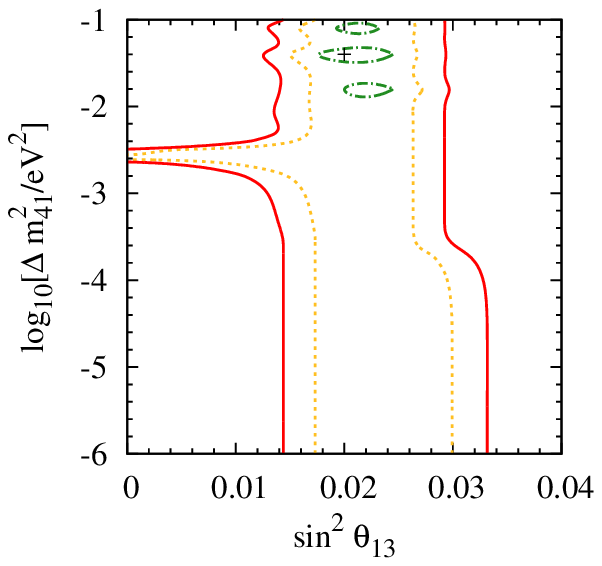}        }
\vspace{-1.2cm}
\caption{\it Confidence level regions at 1$\sigma$, 2$\sigma$, and 3$\sigma$ for 2 dof,
after performing a fit to the Daya Bay data,
in the $(\sin^2 2\theta_{14},\Delta m^2_{41})$ and $(\sin^2 \theta_{13},\Delta m^2_{41})$-planes presented in the
left and right plots, respectively.}
\label{fig:dmth14th13}
\end{center}
\end{figure}

\begin{figure}[]
\begin{center}
\subfigure{%
\hspace{-1.4cm}
\includegraphics[width=0.65\textwidth]{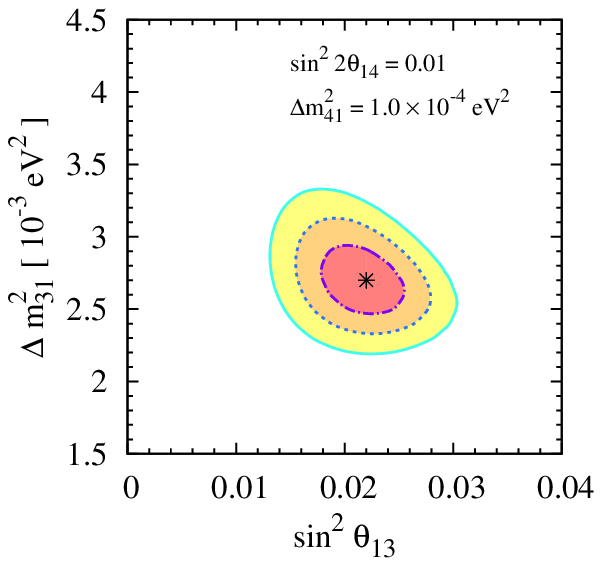}        }%
\subfigure{%
\hspace{-2.4cm}
\includegraphics[width=0.65\textwidth]{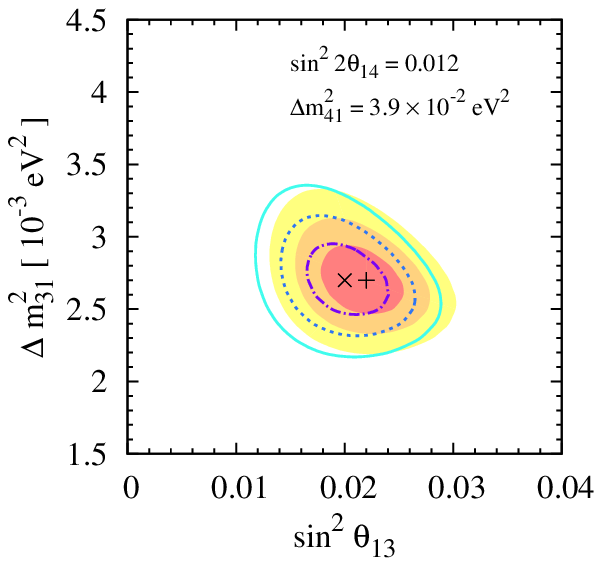}        }
\subfigure{%
\includegraphics[width=0.65\textwidth]{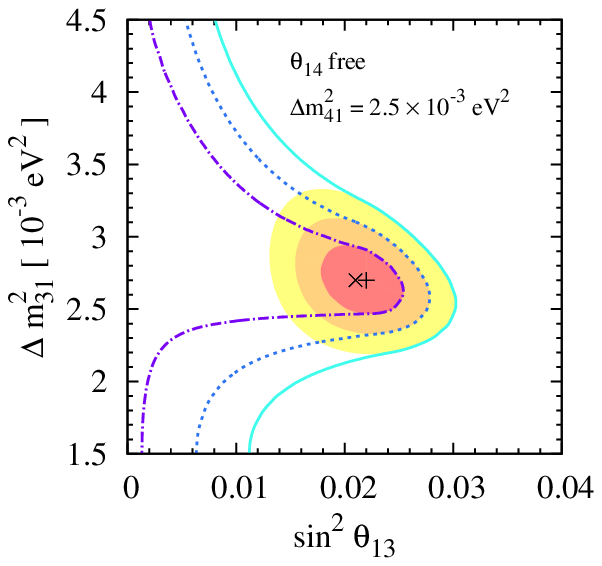}        }
\vspace{-0.3cm}
\caption{\it Confidence level regions at 1$\sigma$, 2$\sigma$, and 3$\sigma$ for 2 dof in the
$(\sin^2 \theta_{13},\Delta m^2_{41} )$-plane after performing a fit to the Daya Bay data.
For the left and the right upper plots, the sterile oscillation parameters are fixed to
$(\sin^2 2\theta_{14},\Delta m^2_{41}) = (10^{-2},10^{-4} \,  {\rm eV^2})$
and $(\sin^2 2\theta_{14},\Delta m^2_{41}) = (0.012,0.039 \, {\rm eV^2})$, respectively.
The lower plot has been obtained fixing
$\Delta m^2_{41} = 2.5 \times 10^{-3} \, {\rm eV^2}$ and varying freely $\theta_{14}$.
}
\label{fig:th13th14dm41th13}
\end{center}
\end{figure}

In particular, we use Gaussian priors defined through the mean value and the 1$\sigma$ error as follows:
\begin{eqnarray}
\sin^2 \theta_{12} & =& 0.306 (1 \pm 5~\%) \;  , \nonumber \\
\sin^2 \theta_{13} &=& 0.021 (1 \pm 20~\%) \;  , \nonumber \\
\Delta m^2_{21} &=& [7.58 (1 \pm 5~\%)]\times 10^{-5}~{\rm eV^2} \; , \nonumber \\
\left| \Delta m^2_{31} \right|  &= & [(2.35 (1 \pm 20~\%)] \times 10^{-3}~{\rm eV^2} \; .
\label{eq:input}
\end{eqnarray}
The central values in Eq.~\eqref{eq:input} have been obtained from Ref.~\cite{Fogli:2011qn}, although  with 1$\sigma$ errors slightly
larger to account for possible (unevaluated) effects due to the presence of sterile neutrinos.
The new parameters $\theta_{14}$ and $\Delta m^2_{41}$ are considered as free parameters:
the mass-squared difference is completely unconstrained in the range $(10^{-6},1)$ eV$^2$,
while for the mixing angle we only considered the upper bound
$\theta_{14} < 20^{\circ}$.
In all figures the green dotted-dashed, yellow dotted, and red solid curves refer to 1$\sigma$, 2$\sigma$, and 3$\sigma$ regions in 2 degrees of freedom (dof), respectively.
The results in the $(\sin^2 2\theta_{14},\Delta m^2_{41})$-plane is shown in the left plot of
Fig.~\ref{fig:dmth14th13} after a marginalization over all the standard oscillation
parameters using the priors defined in Eq.~(\ref{eq:input}), in which we can clearly see that at the smallest confidence level
a best fit point emerges at $(\sin^2 2\theta_{14},\Delta m^2_{41})  = (0.012 , 0.039 \, {\rm eV^2})$.
However, since a relatively large part of the parameter space is still allowed at 2$\sigma$,
it is interesting to analyze the impact of the presence of a third independent mass-squared difference $\Delta m^2_{41}$ on the measurement of $\theta_{13}$.
This is shown
in the right plot of Fig.~\ref{fig:dmth14th13}, obtained after marginalizing over the undisplayed $\theta_{14}$
(limited by $\theta_{14} < 20^{\circ}$) and the other
standard parameters with priors as in  Eq.~(\ref{eq:input}).
We can easily recognize the presence of two distinct regions. One for $\Delta m^2_{41}
\lesssim 10^{-3} \, {\rm eV^2}$ and $\Delta m^2_{41}\gtrsim 5 \times 10^{-3}  \, {\rm eV^2}$ (at 3$\sigma$)
where, as also outlined in Ref.~\cite{Palazzo:2013bsa}, the measurement of $\theta_{13}$ is quite robust and almost unaffected
by sterile neutrinos. The other for $10^{-3} \,{\rm eV^2}\lesssim \Delta m^2_{41}\lesssim 5 \times  10^{-3} \,{\rm eV^2}$ in which, given the strong
interplay between $\theta_{13}$ and $\theta_{14}$ for $\Delta m^2_{41} \sim \Delta m^2_{31}$
in the oscillation probability, $\theta_{13}$ can also become vanishingly small.
For our purposes, it is enough to study three different cases, shown in Fig.~\ref{fig:th13th14dm41th13}
(obtained  marginalizing over the other
standard parameters and on $\theta_{14}$):
$(\sin^2 2\theta_{14},\Delta m^2_{41}) = (10^{-2},10^{-4} \, {\rm eV^2})$ (upper left plot), $(\sin^2 2\theta_{14},\Delta m^2_{41}) = (0.012,0.039 \, {\rm eV^2})$
(upper right plot, corresponding to the best-fit point shown in the left plot of Fig.~\ref{fig:dmth14th13})
and $\Delta m^2_{41} = 2.5\times 10^{-3}$ eV$^2$ with free $\theta_{14}$ (lower plot).
As can be observed from the left and the right upper plots of Fig.~\ref{fig:th13th14dm41th13}, the presence of sterile neutrinos does not affect significantly the determination of the standard
oscillation parameters $\theta_{13}$ and $\Delta m^2_{31}$ for mass-squared differences away from the region $10^{-3} \,{\rm eV^2}\lesssim \Delta m^2_{41}\lesssim 5 \times 10^{-3}\, {\rm eV^2}$. On the other hand, for a mass-squared difference within this range we observe in the lower plot a much larger spread of the allowed values of $\theta_{13}$ and $\Delta m^2_{31}$,
as a consequence of $\Delta m^2_{41} \approx \Delta m^2_{31}$.
As we have mentioned below Eq.~(7), the existence of a sterile neutrino could mimick the effects of a large $\theta_{13}$ in this case.
The best-fit $\theta_{13}$ and $\Delta m^2_{31}$ are however in consistent with their true values. Concretely, we have the best-fit values ($\sin^2\theta_{13}$, $\Delta m^2_{31}$)= (0.022, $2.7\times 10^{-3}~{\rm eV}^2$), (0.020, $2.7\times 10^{-3}~{\rm eV}^2$) and (0.021, $2.7\times 10^{-3}~{\rm eV}^2$) for the upper left, upper right and lower plots, respectively.

\section{Sensitivity at JUNO}
\label{sec:impact_prec}

The JUNO experiment~\cite{Li:2013zyd} has been designed to determine the neutrino mass hierarchy, i.e., the sign of $\Delta m^2_{31}$, by observing the disappearance of reactor electron antineutrinos at a distance of $52.5~{\rm km}$. With high statistics of one hundred thousand $\bar{\nu}_e$ events in six years and an excellent energy resolution $\gamma = 0.03~{\rm MeV}$, the JUNO setup will also have a very good sensitivity to the other standard neutrino oscillation parameters, in particular to $\theta_{12}$ and $\Delta m^2_{21}$. In this section, we explore the impact of sterile neutrinos with a mass-squared difference $\Delta m^2_{41}$ ranging from $10^{-6}~{\rm eV}^2$ to $10^{-1}~{\rm eV}^2$ on precision measurements of ($\theta_{12}$, $\Delta m^2_{21}$) and ($\theta_{13}$, $\Delta m^2_{31}$), and the determination of the neutrino mass hierarchy at JUNO. Moreover, the JUNO sensitivity to sterile neutrinos will be studied and compared with the constraint from the Daya Bay data presented in Sec.~\ref{sec:DB}.

Following the approach in Ref.~\cite{Ohlsson:2013nna}, we perform our simulations for the JUNO setup by using the GLoBES software~\cite{Huber:2003pm,Huber:2004ka,Huber:2007ji}. The true values of the relevant standard parameters are taken from the latest global-fit analysis of neutrino oscillation experiments~\cite{Capozzi:2013csa}:
\begin{eqnarray}\label{eq:paras}
\sin^2\theta_{12} & = &  0.308 \pm 0.017 \; , \nonumber \\
\sin^2\theta_{13} & = &  0.0234 \pm 0.002 \; ,\nonumber \\
\Delta m^2_{21} & = & (7.54 \pm 0.26) \times 10^{-5}~{\rm eV}^2 \; , \nonumber \\
\left| \Delta m^2_{31} \right| & = & (2.43 \pm 0.06) \times 10^{-3}~{\rm eV}^2 \; ,
\end{eqnarray}
where $1\sigma$ errors are assumed to be Gaussian and will be incorporated into our simulations as priors for the corresponding parameters. It is worth mentioning that the true values and uncertainties in Eq.~(\ref{eq:paras}) have been obtained by including the Daya Bay data~\cite{Capozzi:2013csa}, in contrast to those in Eq.~(\ref{eq:input}). Since JUNO is very sensitive to $(\theta_{12},\Delta m^2_{21})$, the priors on these are not relevant here. However, the prior knowledge on $(\theta_{13},\Delta m^2_{31})$ from existing reactor neutrino experiments, such as Daya Bay, is important and will be taken into account.
\begin{figure}[!t]
\begin{center}
\subfigure{%
\hspace{-1.4cm}
\includegraphics[width=0.64\textwidth]{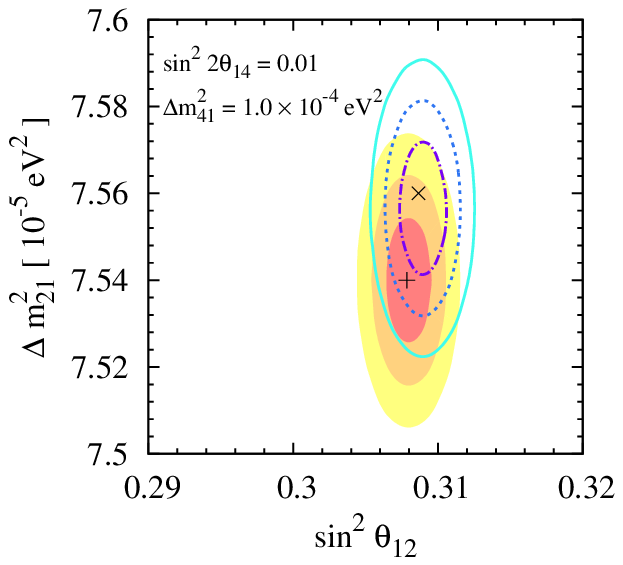}        }%
\subfigure{%
\hspace{-2.4cm}
\includegraphics[width=0.64\textwidth]{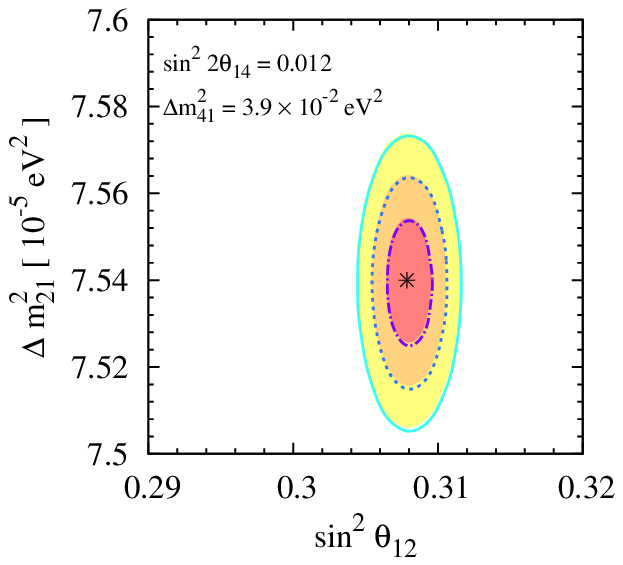}        }
\vspace{-1.2cm}
\caption{\it Illustration for the impact of sterile neutrinos on the experimental sensitivities to $(\sin^2 \theta_{12}, \Delta m^2_{21})$ at JUNO. In our simulations, the true values in Eq.~(\ref{eq:paras}) have been used. The red (dark-gray), orange (gray), and yellow (light-gray) areas stand respectively for the 1$\sigma$, 2$\sigma$, and 3$\sigma$ regions for 2 dof in the case of no sterile neutrinos, while the fit results in the presence of sterile neutrinos are represented by the purple (dotted-dashed), blue (dotted), and cyan (solid) curves. Left plot: For $(\sin^2 2\theta_{14}, \Delta m^2_{41}) = (0.01, 1.0\times 10^{-4}~{\rm eV}^2)$, the best-fit values are $(\sin^2 \theta_{12}, \Delta m^2_{21}) = (0.309, 7.56\times 10^{-5}~{\rm eV}^2)$. Right plot: For $(\sin^2 2\theta_{14}, \Delta m^2_{41}) = (0.012, 3.9\times 10^{-2}~{\rm eV}^2)$, the best-fit values coincide with those in the case of no sterile neutrinos.}
\label{fig:impact_t12_dm21}
\end{center}
\end{figure}
\subsection{The Parameters $\theta_{12}$ and $\Delta m^2_{21}$}
\label{subsec:th12dm12}

In order to illustrate how sterile neutrinos affect the precision measurement of $(\theta_{12}, \Delta m^2_{21})$, we generate neutrino data at JUNO by assuming a light sterile neutrino with $(\sin^2 2\theta_{14}, \Delta m^2_{41}) = (0.01, 1.0\times 10^{-4}~{\rm eV}^2)$. In addition, the true values of the relevant standard parameters are given in Eq.~(\ref{eq:paras}). Then, the generated data are fitted by the standard parameters, with $\theta_{13}$ and $\Delta m^2_{31}$ being marginalized over. As shown in the left plot of Fig.~\ref{fig:impact_t12_dm21}, the best-fit values in this case turn out to be $(\sin^2 \theta_{12}, \Delta m^2_{21}) = (0.309, 7.56\times 10^{-5}~{\rm eV}^2)$ denoted by ``$\times$", which are significantly different from the best-fit values $(\sin^2 \theta_{12},\Delta m^2_{21}) = (0.308, 7.54\times 10^{-5})$ denoted by ``+" in the standard case. The purple dotted-dashed, blue dotted, and cyan solid curves stand for the $1\sigma$, $2\sigma$, and $3\sigma$ contour curves, respectively. The difference between best-fit and true values of $\theta_{12}$ can be well understood from Eq.~(\ref{eq:probapp}), where $\Delta_{42}$ and $\Delta_{41}$ are of the same order of $\Delta_{21}$ and lead to excessive disappearance of reactor antineutrinos. The latter can also be explained by a larger value of $\theta_{12}$, but without sterile neutrinos. On the other hand, for the chosen true values, $|\Delta_{41}| > |\Delta_{21}| > |\Delta_{42}|$ and $\cos^2 \theta_{12} > \sin^2 \theta_{12}$ indicate that sterile neutrinos introduce an additional term of faster oscillations, which can be mimicked by a larger $\Delta m^2_{21}$. However, if the 1$\sigma$ errors of the priors of $\theta_{13}$ and $\Delta m^2_{31}$ are taken of the order of 20\%, the difference between the standard and the nonstandard fits becomes insignificant. For comparison, we present an analysis of JUNO sensitivity in the standard case without sterile neutrinos, and the shaded areas correspond to the $1\sigma$, $2\sigma$, and $3\sigma$ regions, respectively. Given the true values $(\sin^2 2\theta_{14}, \Delta m^2_{41}) = (0.01, 1.0\times 10^{-4}~{\rm eV}^2)$, it is obvious from Fig.~\ref{fig:impact_t12_dm21} that the JUNO sensitivity to $(\theta_{12}, \Delta m^2_{21})$ is essentially not changed, although the best-fit values may deviate from the true values.

If the best-fit values $(\sin^2 2\theta_{14}, \Delta m^2_{41}) = (0.012, 3.9\times 10^{-2}~{\rm eV}^2)$ from Daya Bay data are taken as true values in our simulations, the JUNO sensitivities to $\theta_{12}$ and $\Delta m^2_{21}$ are almost unchanged, as shown in the right plot of Fig.~\ref{fig:impact_t12_dm21}. According to Eq.~(\ref{eq:probapp}), $\Delta m^2_{41} > \Delta m^2_{31} \gg \Delta m^2_{21}$ implies that the contributions from sterile neutrinos can be hidden by the uncertainties of $(\theta_{13}, \Delta m^2_{31})$, in particular for $\sin^2 2\theta_{14} \ll \sin^2 2\theta_{13}$ in our case. For this set of parameters, JUNO is not sensitive enough to place a restrictive constraint.

It is worthwhile to make a comparison between the sensitivity to $(\sin^2 \theta_{12}, \Delta m^2_{21})$ from our simulations and that given by the JUNO Collaboration. In Fig.~\ref{fig:impact_t12_dm21}, the $1\sigma$ error on $\sin^2 \theta_{12}$ is 0.0015 and that on $\Delta m^2_{21}$ is $0.014 \times 10^{-5}~{\rm eV}^2$, corresponding to a precision of $0.49\%$ and $0.19\%$, respectively. In our simulations, only one reactor with thermal power of $35.8~{\rm GW}$ and a flux normalization uncertainty of $3\%$ are considered, and we have ignored the background and other systematics. For the nominal setup and systematic uncertainties considered in Ref.~\cite{Li:2013zyd}, the estimates of the sensitivity to $(\sin^2 \theta_{12}, \Delta m^2_{21})$ from the JUNO Collaboration are $0.54\%$ and $0.24\%$, which are in reasonably good agreement with ours. However, when the bin-to-bin energy uncorrelated uncertainty ($1\%$), the energy linear scale uncertainty ($1\%$), the energy nonlinear uncertainty ($1\%$), and the background ($1\%$) are taken into account, the precisions will be $0.67\%$ and $0.59\%$~\cite{Wang:2014}. Therefore, our simulated sensitivity will be reduced if the background and the above systematic uncertainties are included.

\subsection{The Parameters $\theta_{13}$ and $\Delta m^2_{31}$}
\label{subsec:impact_mass}
\begin{figure}[!t]
\begin{center}
\subfigure{%
\hspace{-1.4cm}
\includegraphics[width=0.64\textwidth]{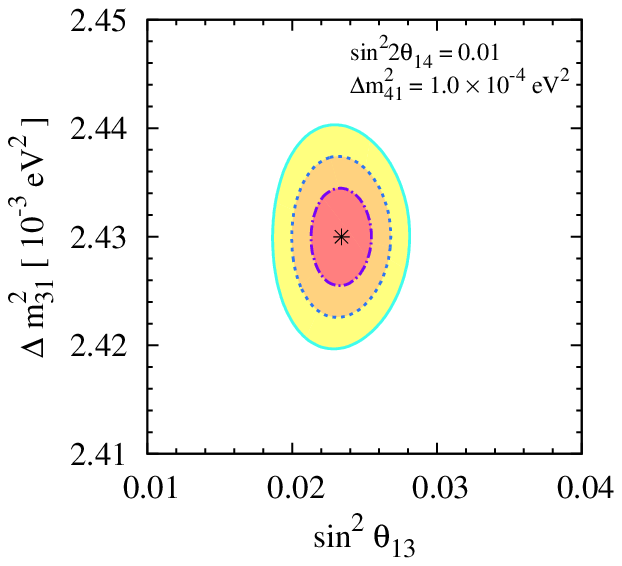}        }%
\subfigure{%
\hspace{-2.4cm}
\includegraphics[width=0.64\textwidth]{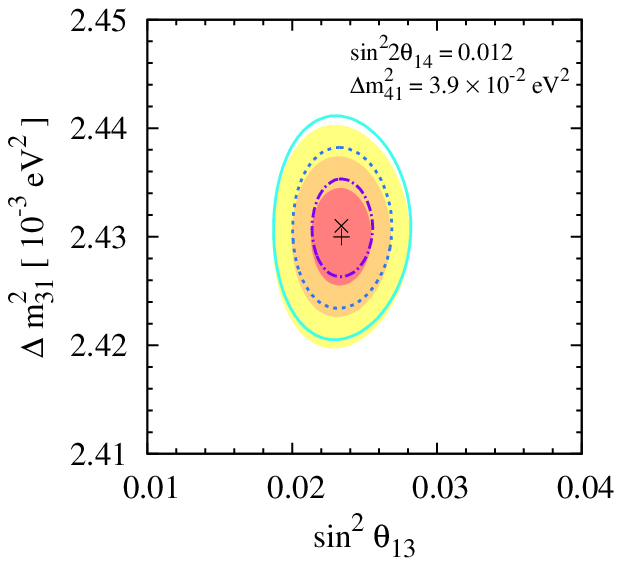}        }
\vspace{-1.2cm}
\caption{\it Illustration for the impact of sterile neutrinos on the experimental sensitivities to $(\sin^2 \theta_{13}, \Delta m^2_{31})$ at JUNO. In our simulations, the true values in Eq.~(\ref{eq:paras}) have been used. The red (dark-gray), orange (gray), and yellow (light-gray) areas stand respectively for the 1$\sigma$, 2$\sigma$, and 3$\sigma$ regions  for 2 dof in the case of no sterile neutrinos, while the fit results in the presence of sterile neutrinos are denoted by the purple (dotted-dashed), blue (dotted), and cyan (solid) curves. Left plot: For $(\sin^2 2\theta_{14}, \Delta m^2_{41}) = (0.01, 1.0\times 10^{-4}~{\rm eV}^2)$, the best-fit values coincide with those in the case of no sterile neutrinos. Right plot: For $(\sin^2 2\theta_{14}, \Delta m^2_{41}) = (0.012, 3.9\times 10^{-2}~{\rm eV}^2)$, the best-fit values deviate slightly from those in the standard case.}
\label{fig:impact_t13_dm31}
\end{center}
\end{figure}

In a similar way, we now consider the impact of sterile neutrinos on the measurement of $(\theta_{13}, \Delta m^2_{31})$ at JUNO. In Fig.~\ref{fig:impact_t13_dm31}, we show the fit of standard parameters to the data generated by oscillation probabilities in the presence of sterile neutrinos. The fit to the data generated with $(\sin^2 2\theta_{14}, \Delta m^2_{41}) = (0.01, 1.0\times 10^{-4}~{\rm eV}^2)$ is given in the left plot, while that with $(\sin^2 2\theta_{14}, \Delta m^2_{41}) = (0.012, 3.9 \times 10^{-2}~{\rm eV}^2)$ in the right plot.

In the former case, the best-fit value of $\theta_{13}$ in the sterile neutrino case coincides exactly with that in the standard case. Moreover, the $1\sigma$, $2\sigma$, and $3\sigma$ contour curves overlap with the edges of shaded regions, which are obtained by generating neutrino data without sterile neutrinos. The reason is two-fold. First, $\Delta_{41} \approx \Delta_{21} \ll \Delta_{31}$ and the corrections to the standard oscillation probability of three active neutrinos can be absorbed into the uncertainties of $(\theta_{12}, \sin^2 \Delta_{21})$. Second, the JUNO setup itself has limited sensitivity to $(\theta_{13}, \Delta m^2_{31})$.

In the latter case, the deviation from the fit without sterile neutrinos is visible, but insignificant. Due to $\Delta_{41} > \Delta_{31}$, the best-fit point is now shifted to a larger value of $\Delta m^2_{31}$. It is now evident that a light sterile neutrino does not affect the measurement of $(\theta_{13}, \Delta m^2_{31})$ at JUNO, which in any event is not very sensitive to these two parameters.

\subsection{The Neutrino Mass Hierarchy}
\begin{figure}[!t]
\begin{center}
\subfigure{
\hspace{-2cm}
\includegraphics[width=1.2\textwidth]{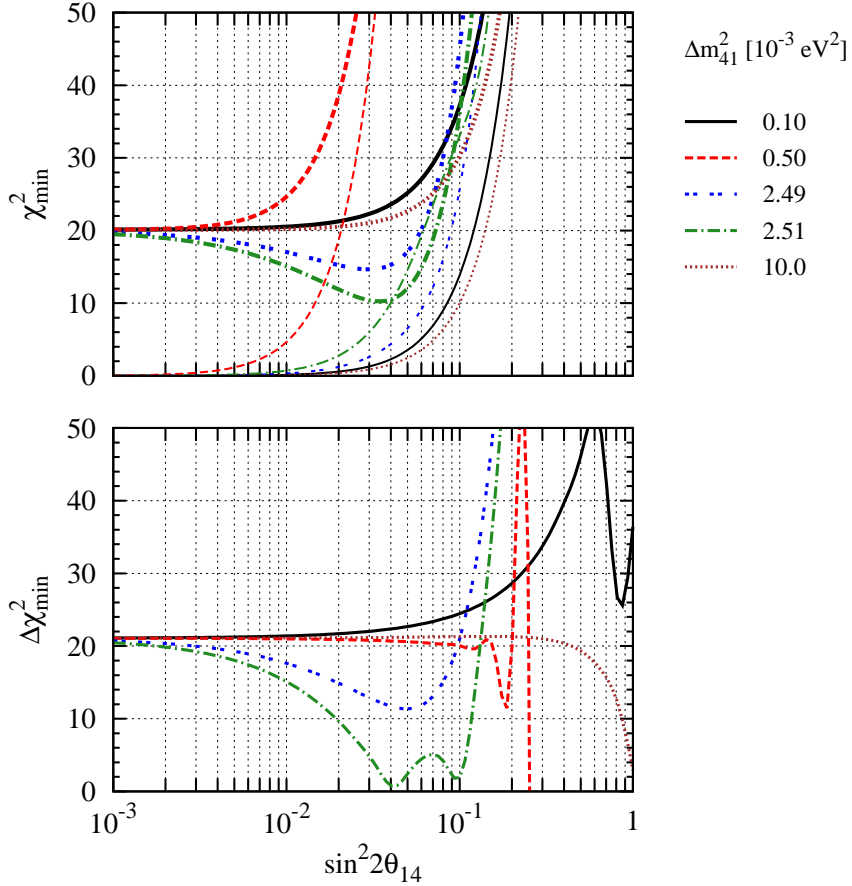} }
\vspace{-2cm}
\caption{\it Impact on the determination of neutrino mass hierarchy at JUNO. In our simulations, neutrino data are generated in the NH case. In the upper plot, thick curves refer to the fits with IH, while the corresponding thin curves to those with NH. In the lower plot, the absolute values of differences between the IH and NH fits $\Delta \chi^2_{\rm min} \equiv |\chi^2_{\rm min} ({\rm IH}) - \chi^2_{\rm min} ({\rm NH})|$ have been given for $\Delta m^2_{41} = 1.0 \times 10^{-4}~{\rm eV}^2$ (solid), $5.0\times 10^{-4}~{\rm eV}^2$ (dashed), $2.49\times 10^{-3}~{\rm eV}^2$ (double-dotted), $2.51\times 10^{-3}~{\rm eV}^2$ (dotted-dashed), and $1.0\times 10^{-2}~{\rm eV}^2$ (dotted).}
\label{fig:impact_mass}
\end{center}
\end{figure}
In Fig.~\ref{fig:impact_mass}, we show the JUNO sensitivity to the neutrino mass hierarchy in the presence of sterile neutrinos. In our simulations, the neutrino data are generated in the NH case and the true values of the standard parameters are given in Eq.~(\ref{eq:paras}). Additionally, the true values of $\Delta m^2_{41}$ are specified in the plot, and the black solid, red dashed, blue double-dotted, green dotted-dashed, and brown dotted curves correspond to $\Delta m^2_{41} = 1.0 \times 10^{-4}~{\rm eV}^2$, $5.0\times 10^{-4}~{\rm eV}^2$, $2.49\times 10^{-3}~{\rm eV}^2$, $2.51\times 10^{-3}~{\rm eV}^2$, and $1.0\times 10^{-2}~{\rm eV}^2$, respectively. The fit to the generated neutrino data has been carried out both in the NH and IH cases. In the upper plot, the values $\chi^2_{\rm min}$ of the IH fits are denoted by thick curves, while those of the NH fits by thin curves of the same kind. The absolute values of the differences between the IH and NH fits, namely $\Delta \chi^2_{\rm min} \equiv |\chi^2_{\rm min} ({\rm IH}) - \chi^2_{\rm min} ({\rm NH})|$, are shown in the lower plot. The value of $\Delta \chi^2_{\rm min}$ can be used to measure the capability of the JUNO setup to discriminate between NH and IH.

It is interesting to observe from the lower plot of Fig.~\ref{fig:impact_mass} that $\Delta \chi^2_{\rm min}$ approximately vanishes at $\sin^2 2\theta_{14} = 0.04$ for $\Delta m^2_{41} = 2.51\times 10^{-3}~{\rm eV}^2$, which corresponds to the green dotted-dashed curve. This can be perfectly understood with the help of Eqs.~(\ref{eq:deltaP}) and (\ref{eq:condition}), where one can see that the oscillation probabilities in the NH and IH cases are equal at this point in parameter space. Therefore, JUNO is unable to pin down the neutrino mass hierarchy in this case. Note that there will be another zero point for $\Delta \chi^2_{\rm min}$ around $\sin^2 2\theta_{14} \approx 0.1$. However, now both $\chi^2_{\rm min}$(IH) and $\chi^2_{\rm min}$(NH) are quite large, implying that three active neutrino oscillations in both the NH and IH cases cannot fit the data well. This indicates that the JUNO setup is sensitive enough to constrain or discover a light sterile neutrino with the corresponding mixing parameters. Except for the mass region $\Delta m^2_{41} \approx\Delta m^2_{31}$, sterile neutrinos have little impact on the determination of the neutrino mass hierarchy.

\subsection{The Sensitivity at JUNO}
\begin{figure}[!t]
\begin{center}
\subfigure{
\includegraphics[width=0.8\textwidth]{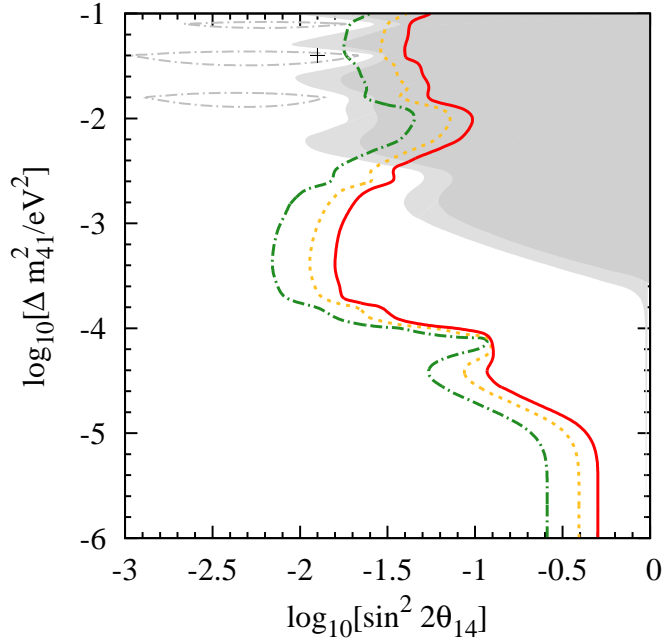} }
\vspace{-0.5cm}
\caption{\it Experimental sensitivity to sterile neutrinos at JUNO. The green (dotted-dashed), yellow (dotted), and red (solid) curves correspond to the 1$\sigma$, 2$\sigma$, and 3$\sigma$ contours  for 2 dof, respectively. For comparison, the fit to Daya Bay data in Fig.~\ref{fig:dmth14th13} has been reproduced, where the dark (light) shaded area is excluded by Daya Bay at the 3$\sigma$ (2$\sigma$) confidence level.}
\label{fig:junosensitivity}
\end{center}
\end{figure}

Finally, let us proceed to explore the sensitivity of the JUNO setup to the mixing parameters of sterile neutrinos. In our simulations, neutrino data are generated by the standard oscillation probabilities and the true values are given in Eq.~(\ref{eq:paras}). The data are fitted by the general oscillation probability with sterile neutrinos, and all the relevant standard oscillation parameters $(\theta_{12}, \Delta m^2_{21})$ and $(\theta_{13}, \Delta m^2_{31})$ are marginalized over. Our results have been depicted in Fig.~\ref{fig:junosensitivity}, and compared with the fit to the Daya Bay data. The dark (light) shaded area is excluded by Daya Bay at the 3$\sigma$ (2$\sigma$) confidence level. At the $3\sigma$ confidence level, compared to the JUNO setup, the Daya Bay experiment has a better sensitivity to sterile neutrinos with $\Delta m^2_{41} \gtrsim 4.0\times 10^{-3}~{\rm eV}^2$. In the low-mass region, i.e., $\Delta m^2_{41} < 4.0\times 10^{-3}~{\rm eV}^2$, JUNO always dominates over Daya Bay in constraining light sterile neutrinos. In this sense, it is therefore clear that reactor neutrino experiments at short and medium baselines are complementary to each other.

The JUNO setup is most sensitive to the mass region from $\Delta m^2_{41} = 10^{-4}~{\rm eV}^2$ to $\Delta m^2_{41} = 10^{-3}~{\rm eV}^2$, where the limit $\sin^2 2\theta_{14} < 10^{-2}$ can be reached. The sensitivity is significantly diminished for $\Delta m^2_{41} \approx \Delta m^2_{21}$. In this case, the oscillation probability in Eq.~(\ref{eq:probexp}) is reduced to the standard one with two independent neutrino mass-squared differences, where the spectral information is not useful in constraining sterile neutrinos.  In the limit of a vanishing $\Delta m^2_{41}$, we obtain $\sin^2 \Delta_{43} \approx \sin^2 \Delta_{31}$ and $\sin^2 \Delta_{42} \approx \sin^2 \Delta_{21}$, implying that the standard neutrino oscillation terms in Eq.~(\ref{eq:probexp}) receive corrections from sterile neutrinos if $\theta_{14}$ is not vanishingly small. Since JUNO has an excellent sensitivity to $\theta_{12}$, it will be able to set an upper bound on $\sin^2 2\theta_{14}$.

It is worthwhile to mention that the experimental constraints on sterile neutrinos exist in the disappearance channel $\bar{\nu}_e \to \bar{\nu}_e$ at reactor neutrino experiments and $\nu_e \to \nu_e$ for solar neutrino experiments. In Ref.~\cite{Kopp:2013vaa}, for $\Delta m^2_{41} \gg 10^{-2}~{\rm eV}^2$, the upper bounds $\sin^2 2\theta_{14} < 0.215$ and $\sin^2 2\theta_{14} < 0.28$ at $95~\%$ confidence level have been derived from long-baseline reactor experiments and from solar plus KamLAND data, respectively. Therefore, our results from the Daya Bay experiment and the future JUNO experiment in Fig.~\ref{fig:junosensitivity} improve the existing bounds in the high-mass region, and provide new constraints in the low-mass region.

\section{Summary and Conclusions}
\label{sec:s&c}

One goal of reactor neutrino experiments is to probe new physics beyond the standard-oscillation paradigm as sub-leading effects in neutrino flavor transitions. Due to high statistical precision and good measurements with the Daya Bay experiment, one can obtain some insight into the hypothesis of sterile neutrinos and put limits on light sterile neutrinos when the active-sterile mass-squared difference is located between $10^{-3}$ and $10^{-1}~{\rm eV}^2$. Restricted by the baseline and energy resolution, the Daya Bay experiment has poor sensitivity to sterile neutrinos with a mass-squared difference below $10^{-3}~{\rm eV}^2$. In contrast, the future JUNO setup features a higher resolution on the neutrino spectrum and has a longer baseline compared to Daya Bay, and hence plays a complementarity role to the current measurements especially in the small mass-squared difference regime. This is particularly relevant for solar neutrinos, since the MSW solution suggests a low energy of the spectra of events at Super-Kamiokande and SNO, which is however not shown in the data. A light sterile neutrino with a mass-squared difference of the order of $10^{-5}~{\rm eV}^2$ and a weak mixing with active neutrinos could explain this suppression \cite{deHolanda:2003tx,deHolanda:2010am}. Furthermore, when the recent detection of B mode polarization from the BICEP2 experiment \cite{Ade:2014xna} is considered, an analysis of the combined CMB data in the framework of LCDM+r models gives $N_{\rm eff}=4.00\pm 0.41$ \cite{Giusarma:2014zza}, which also prefers the existence of extra radiation.

In this work, we have therefore focused on the 3+1 neutrino scenario with only one sterile neutrino and investigated the impact of light sterile neutrinos on short and medium-baseline reactor antineutrino experiments. In particular, we have performed a detail study of antineutrino oscillations and determined the sensitive mass regimes of sterile neutrinos for Daya Bay and JUNO. For both setups, active-sterile neutrino oscillations could in principle mimic the standard oscillations when the active-sterile mass-squared difference is close to one of the standard neutrino mass-squared differences, and hence, one looses sensitivity to sterile neutrinos. Our numerical analysis indicates that the public Daya Bay data suggests an upper limit on the sterile neutrino mixing angle $\sin^22\theta_{14} \lesssim 0.06$ at $3\sigma$ level for the mass-squared difference between $10^{-3}$ and $10^{-1}~{\rm eV}^2$. In addition, for fixed sterile neutrino oscillation parameters, the effects of sterile neutrinos on the determination of $\theta_{13}$ and $\Delta m^2_{31}$ are rather tiny and can be neglected in extracting the standard parameters. Regarding the JUNO setup, the high-energy resolution improves the sensitivity to $\sin^22\theta_{14} \lesssim 0.016$ for $ \Delta m^2_{41} \in (10^{-4} ,10^{-3})~{\rm eV}^2$ and six years of running. However, for a relatively large mass-squared difference, the JUNO sensitivity is not comparable to the one of Daya Bay, due to the longer baseline. When the active-sterile mass-squared difference is around $10^{-4}~{\rm eV}^2$, the measured $\theta_{12}$ and $\Delta m^2_{21}$ deviate from their true values, whereas $\theta_{13}$ and $\Delta m^2_{31}$ are not affected by the sterile neutrino pollution. We have also found a special parameter region that, when $\sin^22\theta_{14} \simeq 0.04$ and $\Delta m^2_{42} \simeq \Delta m^2_{31}$, the sterile neutrino polluted oscillation probability would be almost the same for both NH and IH, indicating that the JUNO setup completely loses its power to discriminate the active neutrino mass hierarchy.

\begin{acknowledgments}
We acknowledge the hospitality and support from the NORDITA scientific program ``News in Neutrino Physics'', April 7--May 2, 2014 during which the initial parts of this study was performed. One of us (I.G.) thanks S. T. Petcov for useful discussions.

This work was supported by the INFN program on ``Astroparticle Physics'' and the European Union FP7-ITN INVISIBLES (Marie Curie Action PITAN-GA-2011-289442-INVISIBLES) (I.G.), MIUR (Italy) under the program Futuro in Ricerca 2010 (RBFR10O36O) (D.M.), the Swedish Research Council (Vetenskapsr{\aa}det), contract no.~621-2011-3985 (T.O.), and the Max Planck Society through the Strategic Innovation Fund in the project MANITOP (H.Z.).
\end{acknowledgments}

\end{document}